\documentclass[12pt]{article}
\pdfoutput=1
\usepackage[utf8]{inputenc}
\usepackage[english]{babel}
\usepackage{tabularx}
\usepackage{array}
\usepackage{graphics}
\usepackage[pdftex]{graphicx}
\usepackage{psfrag}
\usepackage{epsfig}
\usepackage{subfig}
\usepackage{amsmath}
\usepackage{mathtools}
\usepackage{amssymb}
\usepackage{bbold}
\usepackage{bbm}
\usepackage{bm}
\usepackage{setspace}
\usepackage{rotating}
\usepackage{colortbl}
\usepackage{longtable}
\usepackage{slashed}
\usepackage{braket}
\usepackage{lineno}
\makeatletter
\usepackage{textcomp}
\usepackage[usenames,dvipsnames]{xcolor}
\usepackage{relsize}
\usepackage{cite}
\usepackage{float}
\usepackage{simpler-wick}
\usepackage{enumerate}
\usepackage{verbatim}
\usepackage{fancyhdr}
\usepackage{multirow}
\usepackage{arydshln}
\usepackage{enumitem}
\usepackage{hyperref}
\hypersetup{colorlinks,citecolor=nicegreen,linkcolor=niceblue}
\hypersetup{colorlinks=true}
\usepackage{pifont}
\usepackage{physics}

\allowdisplaybreaks

\setlongtables

\newcommand{\bea}{\begin{eqnarray}}
\newcommand{\eea}{\end{eqnarray}}
\newcommand{\beq}{\begin{equation}}
\newcommand{\eeq}{\end{equation}}
\newcommand{\ec}{\end{center}}
\newcommand{\bc}{\begin{center}}

\newcommand{\pdir}{p\kern -5.2pt\raise 0.2ex\hbox {/}}

\newcommand{\vdir}{v\kern -5.75pt\raise 0.15ex\hbox {/}}
\newcommand{\kdir}{k\kern -5.75pt\raise 0.15ex\hbox {/}}
\newcommand{\epsdir}{\epsilon\kern -5.0pt\raise 0.15ex\hbox {/}}
\newcommand{\bvdir}{\bar{v}\kern -5.75pt\raise 0.15ex\hbox {/}}
\newcommand{\Ddir}{D\kern -7.75pt\raise 0.20ex\hbox {/}}
\newcommand{\Adir}{A\kern -7.75pt\raise 0.20ex\hbox {/}}
\newcommand{\ldir}{l\kern -5.0pt\raise 0.2ex\hbox{/}}
\newcommand{\varepsdir}{\varepsilon\kern -5.5pt\raise 0.15ex\hbox{/}}

\newcommand{\nn}{\nonumber}

\makeatother

\newcommand{\s}[1]{\slashed{#1}}

\definecolor{niceblue}{rgb}{0.15,0.15,0.6}
\definecolor{nicegreen}{rgb}{0.1,0.5,0.1}
\definecolor{Red}{rgb}{1.,0.,0.}

\definecolor{Green}{rgb}{0.2,.7,0.2}

\begin{document}
\unitlength = 1mm

\thispagestyle{empty} 

\begin{center}
\vskip 3.4cm\par
{\par\centering \textbf{\Large\bf Revisiting $K \to \pi a$ decays}}
\vskip 1.2cm\par
{\scalebox{.85}{\par\centering \large  
\sc A.~Guerrera$^{a}$ and S.~Rigolin$^{a}$}
{\par\centering \vskip 0.7 cm\par}
{\par\centering \vskip 0.25 cm\par}
{\sl $^a$~Dipartamento di Fisica e Astronomia ``G.~Galilei", Universit\`a degli Studi di Padova e \\
          Istituto Nazionale Fisica Nucleare, Sezione di Padova, I-35131 Padova, Italy} \\
{\vskip 1.65cm\par}}

\end{center}

\vskip 0.85cm
\begin{abstract}
The theoretical calculation for pseudo--scalars hadronic decays $K \to \pi a$ is reviewed. 
While one-loop penguin contributions are usually considered, tree-level processes have most often been overlooked in literature. 
Following the Lepage--Brodsky approach the tree-level contribution to the charged and neutral pseudo--scalar decay in ALP is estimated. 
Assuming generic ALP couplings to SM fermions, the latest NA62/E949 results for the $K^+ \to \pi^+ a$ decay and the present/future KOTO 
results for the  $K^0_L \to \pi^0 a$ decay are used to provide updated bounds on the ALP--fermion Lagrangian sector. Finally, the interplay 
between the tree-level and one-loop contributions is investigated.   
\end{abstract}
\newpage
\setcounter{page}{1}
\setcounter{footnote}{0}
\setcounter{equation}{0}
\noindent

\renewcommand{\thefootnote}{\arabic{footnote}}

\setcounter{footnote}{0}


\newpage

\section{Introduction}
\label{sec:intro}

Light pseudoscalar particles naturally arise in many extensions of the Standard Model (SM) of particle physics. In 
particular they are a common feature of any BSM model endowed with (at least) a global $U(1)$ symmetry spontaneously 
broken at some high scale $f_a \gg v$. Small breaking terms of the global symmetry are then necessary in order to provide 
a mass term, $m_a \ll f_a$, to the generated pseudo Nambu-Goldstone boson (pNGB). Therefore, it may be not inconceivable 
that the first hint of new physics at (or above) the TeV scale could be the discovery of a light pseudoscalar state. 

Sharing a common nature with the QCD axion \cite{Peccei:1977hh,Wilczek:1977pj,Weinberg:1977ma}, these class of pNGBs 
are generically referred to as Axion-Like Particles (ALPs). The key difference between the QCD axion and a generic 
ALP can be summarized in the well-known constraints \cite{Weinberg:1977ma}:
\bea
m_a f_a \approx m_\pi f_\pi 
\label{mafarelation}
\eea
that bounds the QCD axion mass and the $U(1)_{PQ}$ symmetry breaking scale via QCD instanton effects. Present bounds 
on the reference QCD invisible-axion models, like the DNSZ and KVSZ ones \cite{Kim:1979if,Shifman:1979if,Zhitnitsky:1980tq,
Dine:1981rt}, force the axion mass to be typically in the sub-eV range with the symmetry breaking scale $f_a \gtrsim 10^{11}$ 
GeV. In a generic ALP framework, instead, one can assume the ALP mass being determined by some unspecified UV physics, 
besides the usual QCD anomalous contribution, and then relation in Eq.~(\ref{mafarelation}) has not to be enforced. Consequently, 
the ALP mass and the $U(1)_{PQ}$ breaking scale $f_a$ can be taken as independent parameters and in the range phenomenologically 
of interest at present or near future colliders. In this context, a generic ALP can be seen as the generalization of non 
fine-tuned axion models, and permit to look without prejudice to all the possible light pseudo--scalar signatures in Cosmology, 
Astro--Particle and Collider physics.

The ALP parameter space has been intensively explored in several terrestrial facilities, covering a wide energy range~
\cite{Mimasu:2014nea,Jaeckel:2015jla,Bauer:2017ris,Brivio:2017ije,Alonso-Alvarez:2018irt,Harland-Lang:2019zur,
Baldenegro:2018hng,MartinCamalich:2020dfe}, as well as by many astrophysical and cosmological probes~\cite{Cadamuro:2011fd,
Millea:2015qra,DiLuzio:2016sbl}. The synergy of these experimental searches allows to access several orders of magnitude in 
ALP masses and couplings, cf.~e.g.~Ref.~\cite{Irastorza:2018dyq} and references therein. While astrophysics and cosmology impose 
severe constraints on ALPs in the sub-KeV mass range, the most efficient probes of weakly-coupled particles in the MeV-GeV 
range come from experiments acting on the precision frontier~\cite{Essig:2013lka}. Fixed-target facilities such as 
E949~\cite{Artamonov:2009sz,Artamonov:2008qb,Adler:2008zza}, NA62~\cite{CortinaGil:2020fcx,CortinaGil:2021nts} and KOTO~\cite{Ahn:2018mvc} 
and the proposed SHiP \cite{Alekhin:2015byh} and DUNE \cite{Kelly:2020dda} experiments can be very efficient to constrain long-lived 
particles. Furthermore, the rich ongoing research program in the $B$-physics experiments at LHCb~\cite{Aaij:2015tna,Aaij:2016qsm} 
and the $B$-factories~\cite{Masso:1995tw,Bevan:2014iga,Izaguirre:2016dfi,Dolan:2017osp,Kou:2018nap,CidVidal:2018blh,
deNiverville:2018hrc,Gavela:2019wzg,MartinCamalich:2020dfe} offers several possibilities to probe yet unexplored ALP couplings. 

The main goal of this letter, is the detailed analysis of the $K \to \pi a$ decays, in view of the recent NA62 updated 
measurement and the foreseen updates of the KOTO results. While one-loop penguin contributions are usually considered 
\cite{Izaguirre:2016dfi,Dolan:2017osp,Gavela:2019wzg}, the tree-level process contributing to the decay has most often 
been overlooked in literature typically for two reasons. First of all one expects the penguin diagrams to dominate, 
despite the loop suppression, being proportional to the mass of the virtual top quark running in the loop. However, 
in the $K\to \pi a$ case, one has to take into account that the top--loop contribution is CKM suppressed compared to the 
tree-level roughly by a factor $\lambda^4$, and this can partially compensate for the top mass enhancement. The second 
reason is because the tree-level contributions show a much more complicated hadronization structure, being the ALP 
emitted ``inside'' the initial or final meson and so require a dedicated treatment \cite{Lepage:1980fj,Szczepaniak:1990dt}. 
An alternative approach for calculating $K \to \pi a$ decays using chiral perturbation theory can be found, for example, 
in \cite{Georgi:1986df,Bauer:2021wjo}.

The letter is organized as follows. In Sec.~\ref{sec:effective} the effective Lagrangian describing the flavor-conserving 
interactions between the ALP and SM fermions up to dimension five is introduced. In Sec.~\ref{sec:hadronization} the 
tree-level contribution to the $K\to \pi a$ decay is going to be thoroughly discussed. Then, the one-loop penguin 
contributions are shortly reviewed. Finally, in Sec.~\ref{sec:pheno} the interplay between the tree-level and one-loop 
contributions to the $K \to \pi a$ Branching Ratio is thoroughly discussed. Concluding remarks are deferred to 
Sec.~\ref{sec:conclu}.

\section{Effective ALP-SM Fermion Lagrangian} 
\label{sec:effective}

The most general effective Lagrangian describing ALP interactions with SM quarks, including operators up to dimension five, reads:
\bea
\delta \mathcal{L}^a_{\mathrm{eff}} 
 &=&  \frac{\partial_\mu a}{f_a} \left[\overline{U}\, \gamma^\mu \left(C^{(u)}_{L} P_L  \,+\,C^{(u)}_{R} P_R \right) \,U 
                            \,+\,\overline{D}\, \gamma^\mu \left(C^{(d)}_{L} P_L \,+\,C^{(d)}_{R} P_R\right) \,D \right] 
\label{eq:ALP_SMquarks}
\eea
where $f_a$ is the $U(1)_{PQ}$ symmetry breaking scale, $U$ and $D$ the SM up and down flavour triplets and $C_{L,R}$ are general 
$3\times 3$ hermitian matrices. One can, however, heavily reduce the number of independent parameters imposing that the only source 
of flavor--violation in the model arises through the SM Yukawa couplings. Therefore, once the Minimal Flavor Violation (MFV) ansatz 
is assumed, the ALP--quarks Lagrangian becomes
\bea
\delta \mathcal{L}^{a,MFV}_{\mathrm{eff}} &=& 
        -\frac{\partial_\mu a}{ 2 f_a} \sum_{i=quarks} \hspace{-0.25cm} c_i\, \overline{\psi}_i \gamma^\mu \gamma_5 \,\psi_i 
   \,=\, i \frac{a}{f_a} \sum_{i=quarks} \hspace{-0.25cm} c_i m_i \, \overline{\psi}_i \gamma_5 \,\psi_i 
\label{eq:ALP_SMMFV}
\eea
and depends only on six independent flavor diagonal couplings, $c_i$, once the vector fermionic current conservation is implied. Therefore, 
in the MFV framework, flavor--violating ALP couplings can be  generated only at loop level, and is proportional to the SM CKM mixings. 
To further reduce the number of independent parameters, one can additionally assume universal couplings for the up and down sectors, in the 
following denoted as $c_\uparrow$ and $c_\downarrow$, respectively. It may not be unconceivable, in fact, that an hypothetical UV--complete 
model provides different universal PQ charges to the up and down quark sectors \cite{Brivio:2017sdm,Merlo:2017sun,Alonso-Gonzalez:2018vpc}. 
Finally, in the most constrained scenario, one can assume a unique ALP-fermion coupling, often denoted as $c_{a\Phi}$ in the literature, 
as it originates from the dimension five ALP-Higgs operator
\bea
\mathcal{O}_{a\Phi} \equiv i\,\frac{\partial_\mu a}{f_a} \left( \Phi \stackrel{\leftrightarrow}{D}_\mu \Phi\right),
\eea
once the Higgs field is accordingly redefined \cite{Georgi:1986kr}. These additional assumptions will become useful in simplifying 
the phenomenological analysis and their implication will be discussed in Sec.~\ref{sec:pheno}. A general discussion on tree--level 
flavor--violating ALPs couplings to fermions can be found in \cite{MartinCamalich:2020dfe}, while for a recent analysis on 
CP--violating ALP couplings to fermions one is referred, for example, to \cite{DiLuzio:2020oah}.

It might be useful, for simplifying intermediate calculations and explicitly showing the mass dependence of ALP-fermion couplings, to write 
the effective Lagrangian of Eq.~(\ref{eq:ALP_SMMFV}) in the ``Yukawa'' basis instead of the ``derivative'' one. The two versions of the 
effective Lagrangian are equivalent up to operators of $O(1/f_a^2)$. 



%
\section{Meson Hadronization in Flavor Changing Processes} 
\label{sec:hadronization}

Using the effective Lagrangian, implemented with the flavor conserving assumption, of Eq.~\ref{eq:ALP_SMMFV} one can calculate 
the hadronic decay rates of mesons in ALPs. In the following, due to their experimental relevance, the pseudo-scalar meson 
hadronic decays: 
\bea
K^+ \rightarrow \pi^+ \, a \qquad {\rm and} \qquad K^0_L \rightarrow \pi^0 \, a 
\label{invKPidecay}
\eea
will be mainly considered, with the ALP sufficiently long-living to escape the detector without decaying or decaying into invisible 
channels. In such a case the only possible ALP signature is its missing energy/momentum.\footnote{This study will be generalized to 
a wider class of scalar meson decay in a subsequent work \cite{newUS}.} In the following, with $M_{K^{+,0}}, P_{K^{+,0}}$ and 
$M_{\pi^{+,0}}, P_{\pi^{+,0}}$ the mass and 4--momentum of the $K^{+,0}$ and $\pi^{+,0}$ mesons will be denoted, respectively, 
while the ALP mass and 4--momentum will be indicated with $m_a$ and $k_a$.

The processes at hand receive contributions both from tree--level and one--loop diagrams. Typically, for this kind of processes, 
the one--loop penguin diagrams with a top virtual exchange give the dominant contribution, as the $m_t/m_{q,ext}$ ratio largely 
compensate the loop suppression factor. The tree--level and one--loop decay channels involve, however, quite different 
hadronization structures that have to be properly taken into account for correctly comparing the relative contributions to the $K$ 
decay rates. In the following subsections it is briefly illustrated how to handle both contributions in a simple way, applying 
different hadronization methods \cite{Lepage:1980fj,Szczepaniak:1990dt,Carrasco:2016kpy}.

\subsection{The tree-level s-channel process}

\begin{figure}[!ht]
\centering
\includegraphics[scale=0.16]{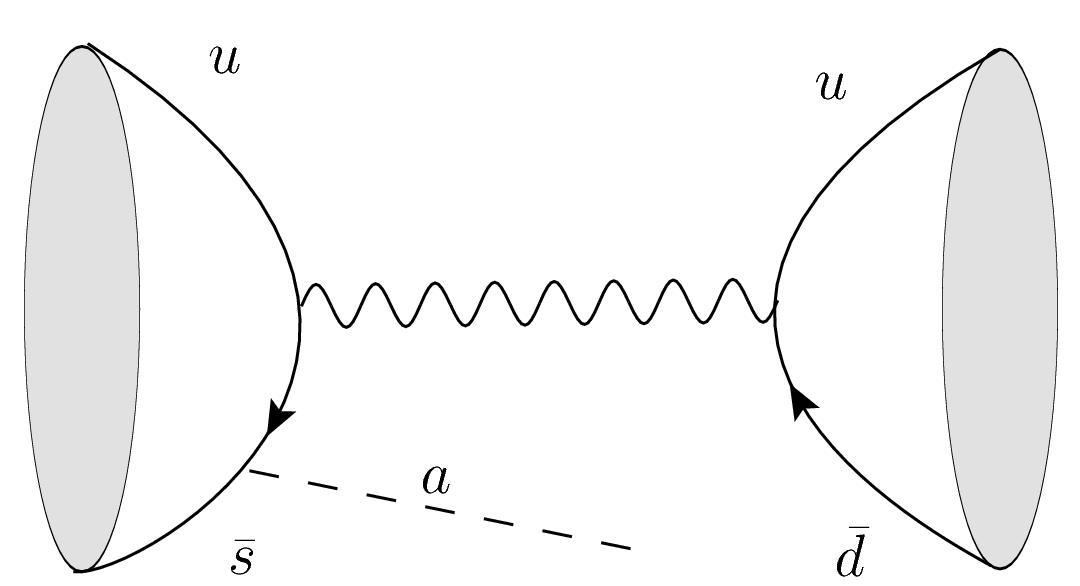}\hspace{2 cm}\includegraphics[scale=0.16]{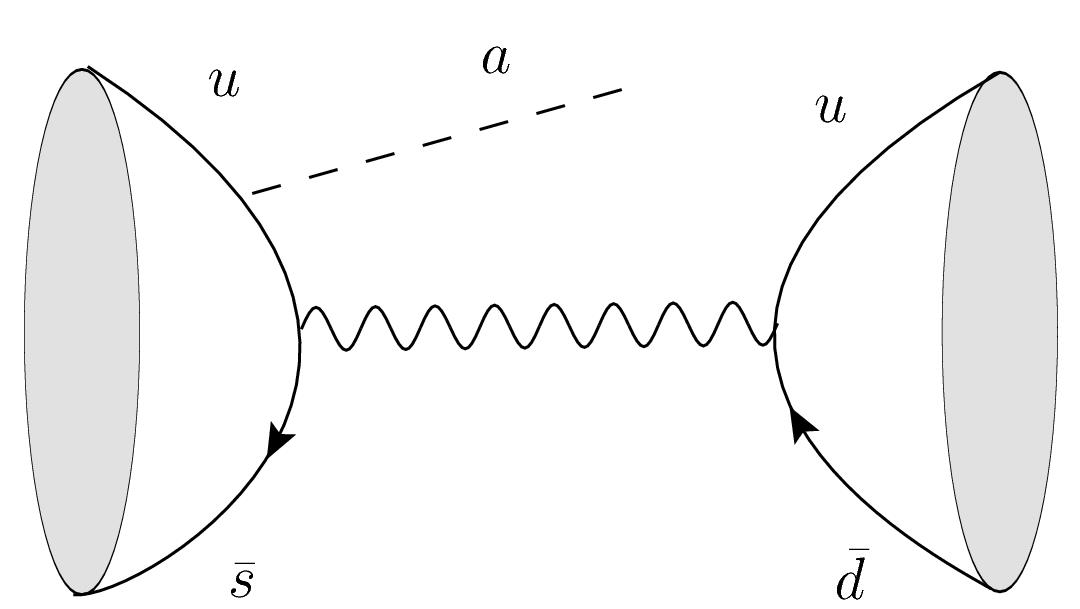}
\caption{Tree level contributions to the $K^+ \rightarrow \pi^+ \, a$ amplitude, with the ALP emitted from the $K^+$ meson. Diagrams 
where the ALP is emitted from the $\pi^+$ quarks are straightforward.
\label{fig:tree_charged}}
\end{figure}

Charged pseudo-scalar meson decays proceed through the s--channel tree-level diagrams of Fig.~\ref{fig:tree_charged}. Here 
only the diagrams where the ALP is emitted from the $K^+$ meson are shown, the ones where the ALP is emitted from the $\pi^+$ follow 
straightforwardly. The tree-level diagram with the ALP emitted from the $W^+$ internal line automatically vanishes, being the 
WW-ALP coupling proportional to the fully antisymmetric 4D tensor.
The hadronization of the s--channel can be done following the Lepage--Brodsky technique \cite{Lepage:1980fj,Szczepaniak:1990dt}.
Leptonic pseudo--scalar decays, $\mathcal{P} \to \ell \,\nu_\ell \, a$, deserve a similar treatment and have been derived in 
\cite{Aditya:2012ay,Hazard:2016fnc}. Let's recall briefly the notation.

The parent meson $K^+$ constituent quarks annihilate into a virtual W boson that then produces the final $\pi^+$ meson partons. The hadronic 
process can be factorized as $\bra{\pi^+}\bar{u}\,\Gamma_{(\pi)}\,d\ket{0}\bra{0}\bar{s}\,\Gamma_{(K)}\,u\ket{K^+}$ with the operator 
insertion $\Gamma_{(\pi)}\otimes\Gamma_{(K)}$ being $\gamma^\mu P_L \otimes \Gamma_{\mu}$ or $\Gamma^{\prime}_{\mu} \otimes \gamma^\mu P_L$ 
depending if the ALP is emitted by the initial or final mesons, with
\bea
\Gamma_\mu \hspace{-0.15cm} &=& \hspace{-0.15cm} \frac{4 G_F}{\sqrt{2}} V^*_{us}V_{ud}
          \left(\frac{c_s\,m_s}{f_a}\,\gamma_5\,\frac{\slashed{k}_a-\slashed{p}_{\bar{s}}+m_s}{m_a^2-2 k_a \cdot p_{\bar{s}}}\,\gamma_\mu P_L 
   \,-\, \frac{c_u\,m_u}{f_a}\,\gamma_\mu P_L\,\frac{\slashed{k}_a-\slashed{p}_u-m_u}{m_a^2 - 2 k_a \cdot p_u}\,\gamma_5 \right) 
          \label{eq:GammaK} \\
\Gamma'_\mu \hspace{-0.15cm} &=& \hspace{-0.15cm}  \frac{4 G_F}{\sqrt{2}} V^*_{us}V_{ud} 
          \left(\frac{c_u\,m_u}{f_a}\,\gamma_5 \,\frac{\slashed{k}_a+\slashed{p}'_u+m_u}{m_a^2-2 k_a\cdot p'_u}\, \gamma_\mu P_L\,
   \,-\, \frac{c_d\,m_d}{f_a}\,\gamma_\mu P_L\,\frac{\slashed{k_a}+\slashed{p}'_{\bar{d}}-m_d}{m_a^2-2 k_a \cdot p'_{\bar{d}}}\,\gamma_5\right).
\label{eq:GammaPi}          
\eea
In Eqs.~(\ref{eq:GammaK}) and (\ref{eq:GammaPi}) with $p_s,p_u$ and $p'_d,p'_u$ the quark momenta of the initial and final meson are denoted, 
respectively. In deriving these Equations the ``Yukawa'' basis for the ALP-SM quark couplings has been explicitly used, providing a simpler 
$\gamma$ structure and showing explicitly the mass dependence of the quark-ALP couplings.

The vector and axial matrix elements can be parameterized in terms of the meson decay constants $f_K$ and $f_\pi$ as: 
\bea
& &\bra{0}\bar{s}\,\gamma^\mu \,\gamma_5 \,u \ket{K^+} = i f_K P_K^\mu \qquad , \qquad \bra{0}\bar{s}\,\gamma^\mu\,u \ket{K^+} = 0 \\
& &\bra{0}\bar{d}\,\gamma^\mu \,\gamma_5 \,u \ket{\pi^+} = i f_\pi P_\pi^\mu \,\,\,\qquad ,\qquad \,\bra{0}\bar{d}\,\gamma^\mu\,u\ket{\pi^+} = 0 \,.
\eea
To compute the $\bra{0} \bar{s}\, \Gamma_\mu \,u\,\ket{K^+}$ and $\bra{\pi^+}\bar{u}\, \Gamma'_\mu\,d\,\ket{0}$ hadronic matrix elements, 
one has to assume a model for describing the effective quark--antiquark distribution inside the meson emitting the ALP. Following 
\cite{Lepage:1980fj,Szczepaniak:1990dt,Aditya:2012ay}, the ground state of a meson $M$ is parameterized with the wave--function
\beq
\Psi_M(x)=\frac{1}{4}\phi_M(x)\gamma^5(\s{P}_M + g_M(x) M_M),
\label{eq:wave_parameter}
\eeq
where ${P}_M$ and $M_M$ denote the momentum and the mass of the meson emitting the ALP\footnote{In the literature the functions $g_M(x)$ 
are conventionally assumed to be constants, $g_M \in [0,1]$.}. In Eq.~(\ref{eq:wave_parameter}), with $x$ one typically denotes the 
fraction of the momentum carried by the heaviest quark in the meson. The function $\phi_M(x)$ describes the meson quarks momenta 
distribution, that for heavy and light mesons reads, respectively:
\bea
\phi_H(x) \propto  \left[\frac{\xi^2}{1-x}+\frac{1}{x}-1\right]^{-2} \qquad , \qquad \phi_L(x) \propto  x(1-x) \,,
\eea
with the normalization fixed such that:
\beq
\int_0^1 dx\,\phi_M(x) = 1.
\label{eq:normalization}
\eeq
The parameter $\xi$ in $\phi_H(x)$ is a small parameter typically of $O(m_q/m_Q)$, being $q$ and $Q$ the light and heavy quark in the 
meson. The mass function $g_M(x)$ is usually taken to be a constant varying from $g_H(x) \approx 1$ and $g_L(x) \ll 1$ for a heavy or a 
light meson, respectively. The hadronic matrix element can then be obtained by integrating over the momentum fraction $x$ the trace of 
the $\Gamma^\mu$ amplitude over the meson wave--function $\Psi_M(x)$:  
\beq
\bra{0} \bar{Q}\,\Gamma^{\mu}\,q \ket{M} \equiv i f_M \int_0^1 dx\,\mathrm{Tr}\left[\Gamma^{\mu} \Psi_M (x)\right]\,.
\label{eq:mesonic_parametrization}
\eeq
In Eqs.~(\ref{eq:wave_parameter}-\ref{eq:mesonic_parametrization}), a slightly different notation with respect to the cited literature 
is used, in particular, as for weak decays colors are left unchanged, all the color matrices/traces have been removed from scratch. 
Moreover, the functions $\Phi_M(x)$ have been normalized to one, in such a way that in Eq.~(\ref{eq:mesonic_parametrization}) the mesonic 
form factor can be explicitly factorized.  

By inserting Eqs.~(\ref{eq:GammaK}-\ref{eq:GammaPi}) into Eq.~(\ref{eq:mesonic_parametrization}), and making the following assignments for 
the initial and final quark momenta:
\bea
p_{\bar{s}}  &=& x P_K \qquad , \qquad p_u  = (1-x) P_K \nn \\
p'_{\bar{d}} &=& x P_\pi \qquad \,, \, \qquad p'_u = (1-x) P_\pi \nn 
\eea
one obtains the following decay amplitudes for the $K^+$-ALP and $\pi^+$-ALP emission processes:
\bea
\mathcal{M}_{K^+} &=& \frac{G_F}{\sqrt{2}} (V^*_{us} V_{ud})\,f_K\, f_\pi \, (k_a\cdot P_\pi) \frac{M_K}{f_a} \,\times\, \label{eq:MKALP} \\ 
     & & \hspace{0.75cm} \times \int^1_0 \left\{ \frac{c_s\,m_s\,\theta(x-\delta^K_a)}{m_a^2-2\,x\,k_a\cdot P_K} - 
         \frac{c_u\,m_u\,\theta(1-x-\delta^K_a)}{m_a^2-2\,(1-x)\,k_a\cdot P_K} \right\} \,\phi_K(x)\,g_K(x)\, dx \nn \\
\mathcal{M}_{\pi^+} &=& \frac{G_F}{\sqrt{2}} (V^*_{us} V_{ud})\,f_K\, f_\pi\, (k_a\cdot P_K) \frac{M_\pi}{f_a} \,\times\, \label{eq:MPiALP}\\
     & & \hspace{0.75cm} \times \int^1_0 \left\{ \frac{c_d\,m_d\,\theta(x-\delta^\pi_a)}{m_a^2-2\,x\,k_a\cdot P_\pi} - 
        \frac{c_u\,m_u\,\theta(1-x-\delta^\pi_a)}{m_a^2-2\,(1-x)\,k_a\cdot P_\pi} \right\} \,\phi_\pi(x)\,g_\pi(x)\,dx \nn
\eea
with $\delta^M_a = m_a^2/(2 k\cdot P_M)$ an explicit cutoff introduced in the fractional momentum to remove the unphysical singularities 
appearing in the integrals. 
The result in Eq.~(\ref{eq:MKALP}) is in agreement with the decay amplitude for $B^\pm \to \ell \,\bar{\nu}_\ell \, a$ calculated in 
\cite{Aditya:2012ay} once $c_u=c_s=1$ is assumed, the pion hadronic current is replaced by the leptonic one, and K quantities replaced by 
the corresponding B meson ones. Eqs.~(\ref{eq:MKALP}) and (\ref{eq:MPiALP}) represent the main result of this section.

Few comments are in order. To numerically evaluate the $K^+\to\pi^+\, a$ branching ratio one has to assume a specific form of the hadronic 
functions $\phi_K(x)$, $g_K(x)$, $\phi_\pi(x)$ and $g_\pi(x)$, and assign a low energy meaning to the quark masses. This, inevitably, 
introduces some model dependence in the calculation. To have an order of magnitude estimate of the $\mathcal{M}_{K^+}$ amplitude, one 
can consider the two extreme cases and treat the K-meson either as a light meson (i.e. assuming an exact global $SU(3)$ symmetry) or as 
an heavy meson (i.e. $m_s \gg m_u,m_d$). In the light meson approximation, substituting the light quarks with the corresponding partons, 
i.e. $\hat{m}_s=\hat{m}_u=M_K/2$, one obtains, for a massless ALP:
\bea
\mathcal{M}^L_{K^+} \approx - \frac{3\, G_F \,f_K\,f_\pi}{4\sqrt{2}} (V^*_{us} V_{ud}) \frac{M^2_K}{f_a}\,g_K\,\left(c_s-c_u\right) \,.
\label{eq:MKLALPapprox}
\eea
Conversely, in the heavy meson approximation, one can assume $\hat{m}_u = \xi M_K$ and $\hat{m}_s = (1-\xi) M_K$, with $\xi = m_u/m_s$. 
Moreover, approximating $\phi_K(x) \approx \delta(1-x-\xi)$ as suggested in \cite{Bhattacharya:2018msv}, one obtains, in the $m_a=0$ limit:
\bea
\mathcal{M}^H_{K^+} \approx  - \frac{G_F \,f_K\,f_\pi}{2\sqrt{2}} (V^*_{us} V_{ud}) \frac{M^2_K}{f_a}\,g_K\,\left(c_s-c_u\right) \,.
\label{eq:MKHALPapprox}
\eea
From the approximate formulas of Eq.~(\ref{eq:MKLALPapprox}) and (\ref{eq:MKHALPapprox}) one can estimate roughly the order of magnitude 
of the uncertainties introduced in the calculation by the hadronization procedure for the K meson. To reproduce the numerical results of 
the following section, an intermediate approach will be, instead, considered: the heavy meson function will be used, with the two partons 
defined as $\hat{m}_u = m_u + \Lambda$ and $\hat{m}_s = m_s +\Lambda$ with $\Lambda = (M_K-m_u-m_s)/2$ a parameter of order $\Lambda_{QCD}$. 
Conversely, for the estimation of the $\mathcal{M}_\pi$ amplitude, one can safely assume to parametrize the pion using the light meson 
wave-function. If one take $g_\pi(x) \approx 0$, as customarily suggested in the literature the pion contribution automatically vanishes. 
A conservative estimate can however be obtained by setting, for example, $g_\pi/g_K \approx M_\pi/M_K$, which predicts the following upper 
bound to the ratio
\bea
R_{\pi K} = \left| \frac{\mathcal{M}_{\pi^+}}{\mathcal{M}_{K^+}} \right| \lesssim \left(\frac{M_\pi}{M_K}\right)^3 \simeq 1.\times 10^{-2} \nn \,.
\label{MpiKapprox}
\eea
For this reason, even in the numerical calculation one can neglect the ALP-$\pi$ emission as expected on a general ground, once same 
order ALP couplings to $u,d$ and $s$ quarks are assumed. 

Finally, from Eqs.~(\ref{eq:MKLALPapprox}) and (\ref{eq:MKHALPapprox}) it appears evident the presence of an ``accidental'' 
cancelation if $c_s=c_u$ is assumed. This cancellation is still partially at work even when the full $\phi_K(x)$ is used and indicates a 
possible underestimation of the $\mathcal{M}_K$ amplitude (and consequently on the ALP-quark coupling limits) in a ``universal'' ALP--SM 
quark coupling scenario compared to the general case. 

\subsection{The tree-level t-channel process}

\begin{figure}[!ht]
\centering
\includegraphics[scale=0.16]{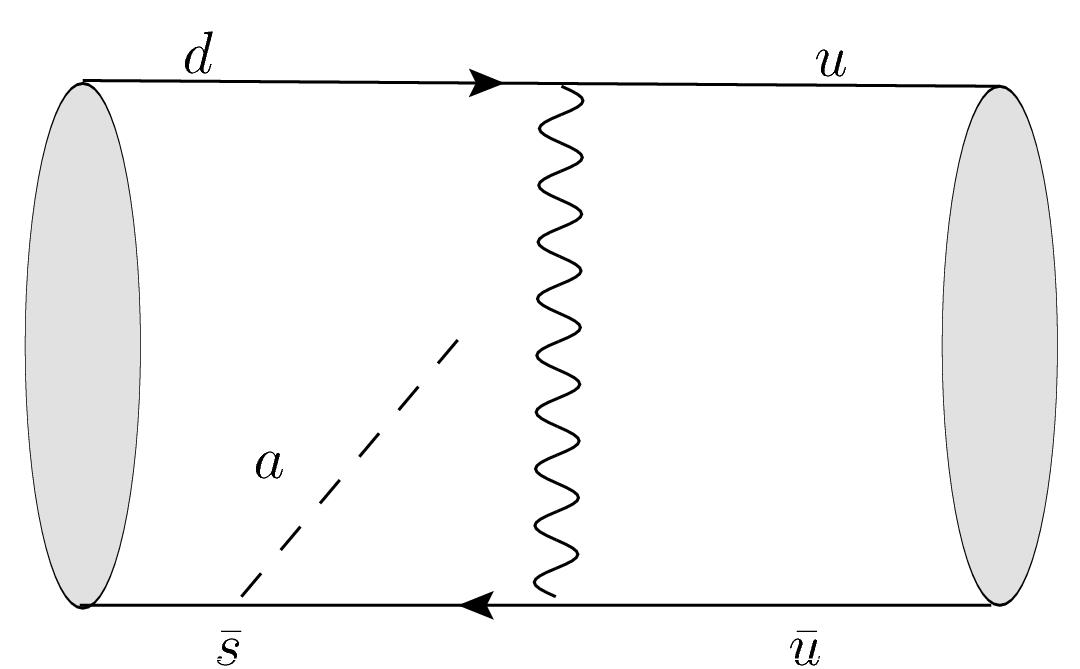}\hspace{2 cm}\includegraphics[scale=0.16]{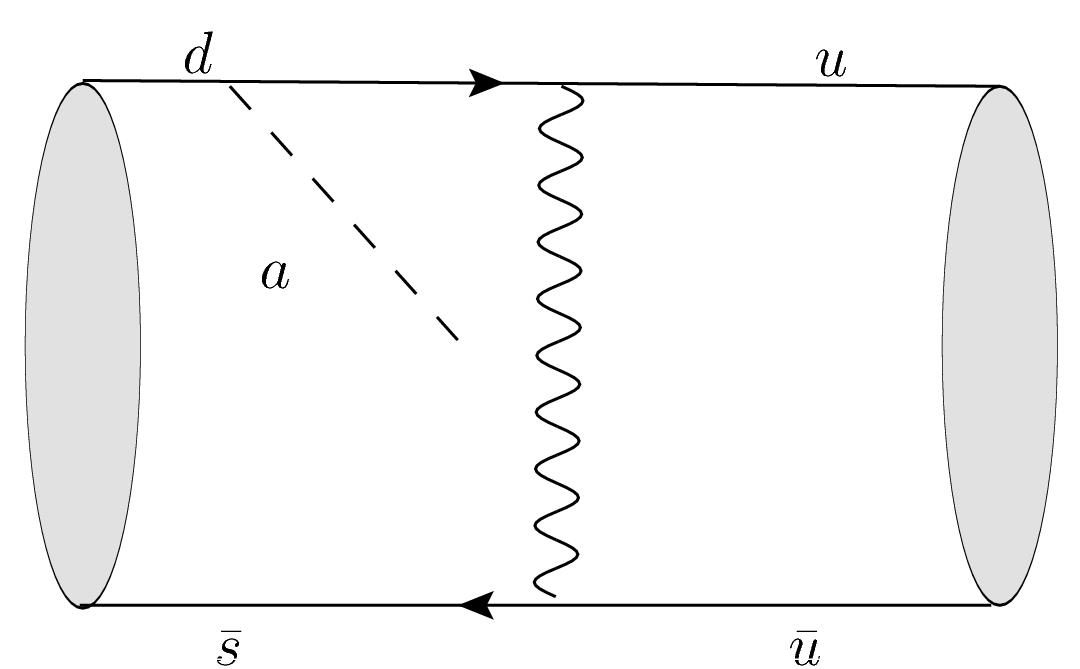}
\caption{Tree level contribution to the amplitude for the $K^0 \rightarrow \pi^0 \, a$ decay, with the ALP emitted from the $K^0$ 
meson. Diagrams where the ALP is emitted from the $\pi^0$ quarks are straightforward. Similar diagrams can be depicted for the 
CP conjugate process $\bar{K}^0 \rightarrow \pi^0 \, a$.
\label{fig:tree_neutral}}
\end{figure}

Neutral pseudo-scalar meson decays proceed through the t--channel tree-level diagrams of Fig.~\ref{fig:tree_neutral}. Here only the 
diagrams where the ALP is emitted from the $K^0$ meson are shown, the ones where the ALP is emitted from the $\pi^0$ follow 
straightforwardly. The tree-level diagram with the ALP emitted from the $W^+$ internal line automatically vanishes, as for the 
s--channel case. The hadronization of the t--channel can be done along the lines depicted in \cite{Lepage:1980fj,Szczepaniak:1990dt}. 

\noindent 
Using the conventions defined in Eqs.~(\ref{eq:wave_parameter}) and (\ref{eq:normalization}), the neutral hadronic matrix element for the 
$K^0$ transition reads:
\beq
\bra{\pi^0}\,\Gamma_1 \otimes \Gamma_2 \ket{K^0} \equiv 
   - \frac{f_K f_\pi}{\sqrt{2}} \int_0^1 dx \int_0^1 dy\,\mathrm{Tr}\left[\Psi_\pi(y)\Gamma_{(1)}\Psi_K(x)\Gamma_{(2)} \right]\,,
\label{eq:tchannel_parametrization}
\eeq
with the $1/\sqrt{2}$ factor taking care of the Clebsch--Gordon suppression of the decay into the neutral $\pi$ meson.

The operator insertion $\Gamma_{(1)}\otimes\Gamma_{(2)}$ is defined as $\gamma_\mu P_L \otimes \Gamma^{\mu}_{(\bar{q})}$ or 
$\Gamma^{\mu}_{(q)} \otimes \gamma_\mu P_L$ depending if the ALP is emitted from the $\bar{q}$ anti--quark (i.e. $\bar{s}$ for the 
$K^0$ and $\bar{u}$ for the $\pi^0$ meson) or from quark $q$ (i.e. the $d$ quark for the $K^0$ and the $u$ quark for the $\pi^0$ meson), 
respectively, with
\bea
\Gamma^\mu_{(\bar{q})}\hspace{-0.15cm} &=& \hspace{-0.15cm} - \frac{4 G_F}{\sqrt{2}} V^*_{us}V_{ud}
     \left(\frac{c_s\,m_s}{f_a}\,\gamma_5\,\frac{\slashed{k}_a-\slashed{p}_{\bar{s}}+m_s}{m_a^2-2 k_a \cdot p_{\bar{s}}}\,\gamma_\mu P_L \,-\,
     \frac{c_u\,m_u}{f_a}\,\gamma_\mu P_L\,\frac{\slashed{k}_a+\slashed{p}'_{\bar{u}}-m_u}{m_a^2-2 k_a\cdot p'_{\bar{u}}}\,\gamma_5\right) 
     \hspace{0.4cm} 
\label{eq:Gamma_qb} \\
\Gamma^\mu_{(q)} \hspace{-0.15cm} &=& \hspace{-0.15cm} - \frac{4 G_F}{\sqrt{2}} V^*_{us}V_{ud} 
          \left(\frac{c_d\,m_d}{f_a}\,\gamma_\mu P_L\,\frac{\slashed{p}_d-\slashed{k}_a+m_d}{m_a^2-2 k_a\cdot p_d}\,\gamma_5 \,+\,
          \frac{c_u\,m_u}{f_a}\,\gamma_5\,\frac{\slashed{k}_a+\slashed{p}'_u+m_u}{m_a^2-2 k_a \cdot p'_u}\,\gamma_\mu P_L\right)
\label{eq:Gamma_q}          
\eea
The $\bar{K}^0\to\pi^0\,a$ decay amplitude can be obtained similarly: 
\beq
\bra{\pi^0}\,\bar{\Gamma}_1 \otimes \bar{\Gamma}_2 \ket{\bar{K}^0} \equiv 
   -\frac{f_K f_\pi}{\sqrt{2}} \int_0^1 dx \int_0^1 dy\,\mathrm{Tr}\left[\Psi_\pi(y)\bar{\Gamma}_{(1)}\Psi_K(x)\bar{\Gamma}_{(2)} \right]\,,
\label{eq:tchannel_bparametrization}
\eeq
with the operator insertions $\bar{\Gamma}_{(1)}\otimes\bar{\Gamma}_{(2)}$ being $\gamma_\mu P_L \otimes \bar{\Gamma}^{\mu}_{(q)}$ or 
$\bar{\Gamma}^{\mu}_{(\bar{q})} \otimes \gamma_\mu P_L$ with
\bea
\bar{\Gamma}^\mu_{(q)}\hspace{-0.2cm} &=& \hspace{-0.15cm} - \frac{4 G_F}{\sqrt{2}} V_{us}V^*_{ud}
     \left(\frac{c_s\,m_s}{f_a}\,\gamma_5\,\frac{\slashed{p}_s-\slashed{k}_a+m_s}{m_a^2-2 k_a \cdot p_s}\,\gamma_\mu P_L \,+\,
     \frac{c_u\,m_u}{f_a}\,\gamma_\mu P_L\,\frac{\slashed{k}_a+\slashed{p}'_u+m_u}{m_a^2-2 k_a\cdot p'_u}\,\gamma_5\right) 
     \hspace{0.5cm}  \label{eq:bGamma_qb} \\
\bar{\Gamma}^\mu_{(\bar{q})} \hspace{-0.2cm} &=& \hspace{-0.15cm} - \frac{4 G_F}{\sqrt{2}} V_{us}V^*_{ud} 
     \left(\frac{c_d\,m_d}{f_a}\,\gamma_\mu P_L\,\frac{\slashed{k}_a-\slashed{p}_{\bar{d}}+m_d}{m_a^2-2 k_a\cdot p_{\bar{d}}}\,\gamma_5\,-\,
     \frac{c_u\,m_u}{f_a}\,\gamma_5\,\frac{\slashed{k}_a+\slashed{p}'_{\bar{u}}-m_u}{m_a^2-2 k_a \cdot p'_{\bar{u}}}\,\gamma_\mu P_L\right)
                     \label{eq:bGamma_q}          
\eea
Adopting the same phase conventions as \cite{Buras:1998raa}, one defines the neutral Kaon mass eigenstates:
\bea
K^0_L &=& \frac{1}{\sqrt{2(1+|\tilde{\epsilon}|^2)}}\left( \left(1+\tilde{\epsilon}\right) K^0 + \left(1-\tilde{\epsilon}\right) \bar{K}^0\right) \\
K^0_S &=& \frac{1}{\sqrt{2(1+|\tilde{\epsilon}|^2)}}\left( \left(1+\tilde{\epsilon}\right) K^0 - \left(1-\tilde{\epsilon}\right) \bar{K}^0\right).
\eea
By making the following assignments for the initial and final quark momenta,
\bea
p_{\bar{s}}  &=& x P_K \qquad , \qquad p'_u  = (1-x) P_K \nn \\
p_d &=& y P_\pi \qquad \, , \,\qquad p'_{\bar{u}} = (1-y) P_\pi \nn 
\eea
the amplitude for the $K^0_L \to \pi^0\,a$ decay, when the ALP emitted by the $K^0_L$ meson, reads: 
\bea
\mathcal{M}_{K^0_L} &=&
  -\frac{ \tilde{\epsilon}\,G_F}{2 \sqrt{2}}\,\Re[V^*_{us} V_{ud}]\,f_{\bar{K}}\, f_{\pi} \, (k_a\cdot P_\pi) \frac{M_K}{f_a} \,\times\, 
    \hspace{0.5cm} \nn \\ 
 & & \hspace{-0.75cm} \times \int^1_0 \left\{ \frac{c_s\,m_s\,\theta(x-\delta^K_a)}{m_a^2-2\,x\,k_a\cdot P_K} - 
    \frac{c_d\,m_d\,\theta(1-x-\delta^K_a)}{m_a^2-2\,(1-x)\,k_a\cdot P_K} \right\} \,\phi_K(x)\,g_K(x)\, dx 
\label{eq:MK0LALP}
\eea
once the trivial integration in $y$ is performed. As one can notice from Eq.~(\ref{eq:MK0LALP}), the amplitude for the $K^0_L$ decay is proportional 
to the oscillation CP violation parameter $\tilde{\epsilon}$, as expected from general considerations on the CP properties of $K^0$ and $\bar{K}^0$ 
decays, and from the absence of imaginary part in the CKM for the tree--level diagram, i.e. $\Im[V^*_{us} V_{ud}]=0$. Consequently the 
$K^0_L$ decay amplitude is suppressed by $O(10^{-3})$, with respect to the corresponding tree--level charged process. Conversely, the $K^0_S 
\to \pi^0\, a$ decay amplitude can be obtained from the result of Eq.~\ref{eq:MK0LALP}, simply removing the $\tilde{\epsilon}$ suppression factor, 
and would be of the same order as the charged process.

The amplitude contribution when the ALP is emitted from the $\pi^0$ meson is not explicitly reported here, showing the same $M_\pi/M_K$ 
suppression as for the charged decay case.

\subsection{The one-loop process}

\begin{figure}[!th]\center
\includegraphics[scale=0.15]{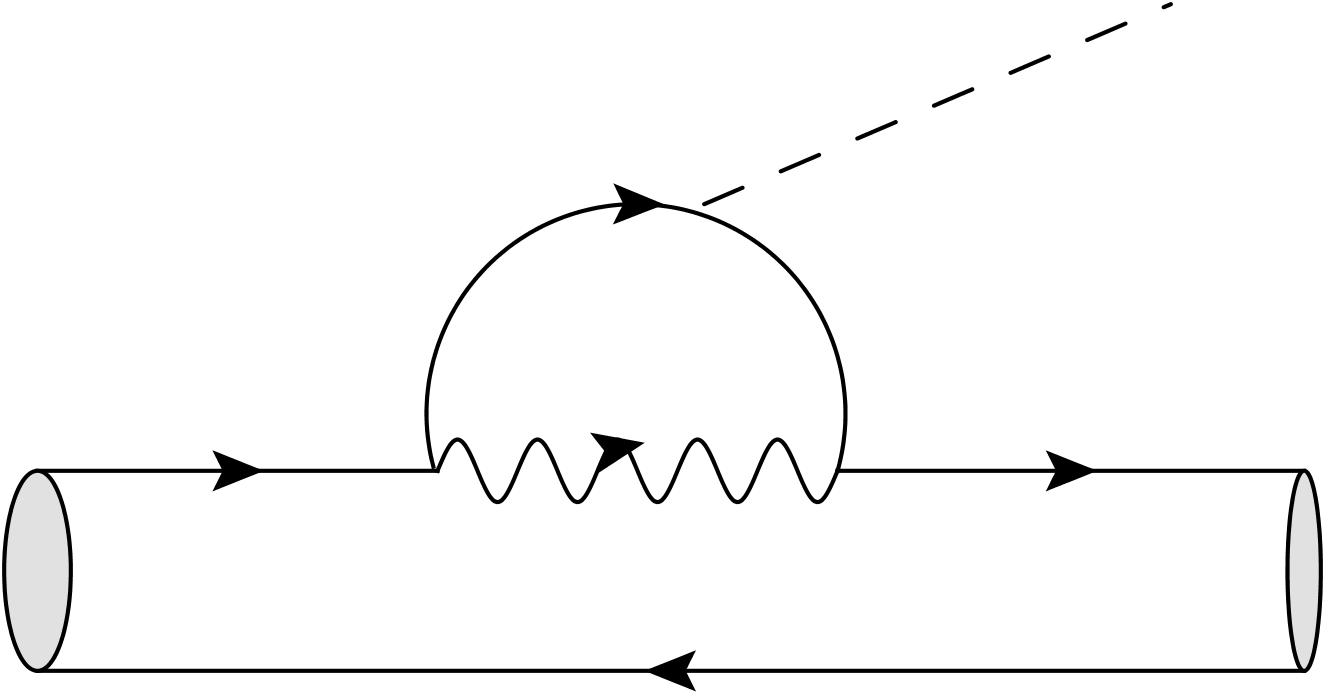}\hspace{1 cm}\includegraphics[scale=0.15]{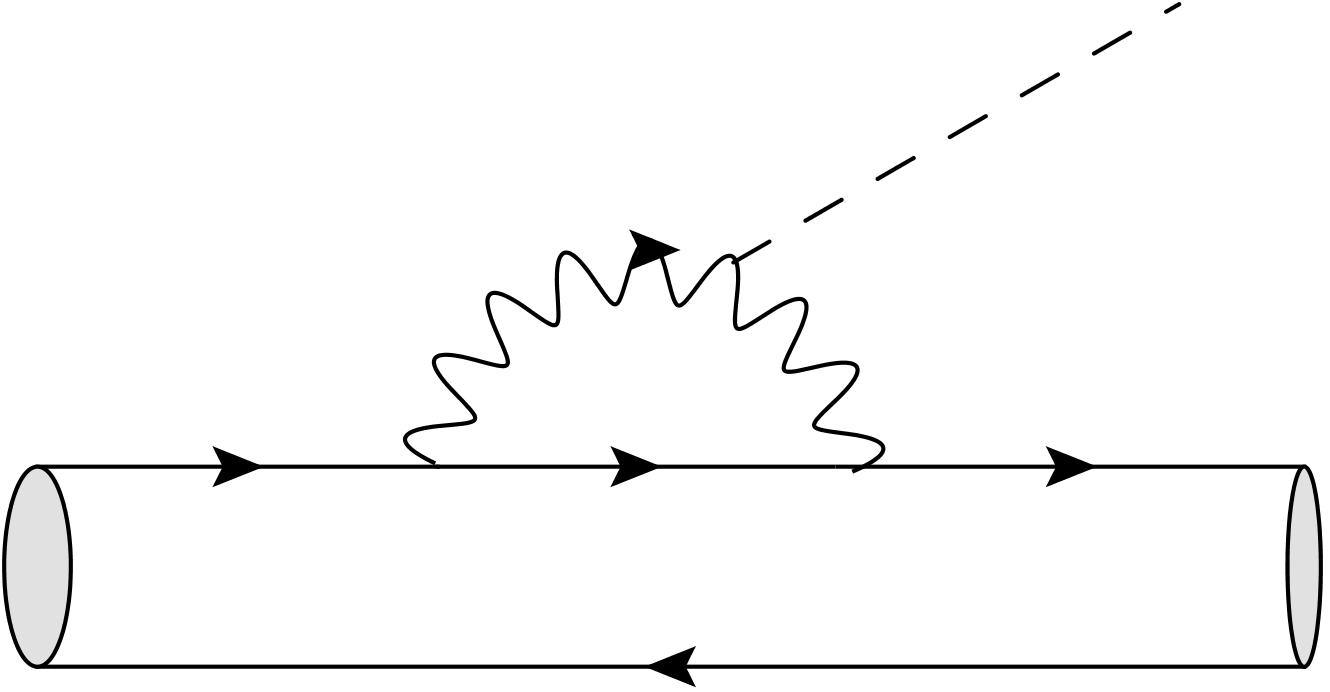}
\caption{One-loop penguin contributions \label{fig:figure_loop}}
\end{figure}

Charged and neutral pseudo-scalar $K \to \pi \,a$ meson decays, assuming flavour conserving fermion-ALP interactions of Eq.~(\ref{eq:ALP_SMMFV}),
receive contributions at one--loop level\cite{Izaguirre:2016dfi,Bauer:2017ris,Gavela:2019wzg} from the diagrams shown in 
Fig.~\ref{fig:figure_loop}. In the following only the contribution arising from fermion-ALP interaction will be considered.

In this kind of processes, only one quark line participate to the ALP emission, the other quark being a spectator. Customarily 
the hadronization of a matrix element between two pseudo-scalar meson mediated by a vector current, where one of the quark does 
not interact can be factorised as
\bea
\bra{P}\bar{q}_1\gamma^\mu Q_2\ket{M}=f_+(q^2)(P_M+P_P)^\mu+f_0(q^2)\,q^\mu
\label{loop_formfactors}
\eea
with $q=P_M-P_P$. The form factors $f_{+,0}(0)=1$ in the isospin symmetric limit, while the non approximated, $q^2$ dependent, form factors 
are obtained from LQCD calculation \cite{Carrasco:2016kpy}. From Eq.~(\ref{loop_formfactors}) the amplitude for the $K^+ \to \pi^+ \,a$ 
decay reads:
\beq
{\mathcal M}^{L}_{K^+} = \frac{G_F\, m^2_t}{4\sqrt{2}\pi^2} (V_{ts}V_{td}^*) \frac{M^2_{K^+}}{f_a}\left(1-\frac{M^2_{\pi^+}}{M^2_{K^+}}\right)
                           \! \left[f_+(m_a^2) + \frac{m_a^2}{M^2_{K^+}-M^2_{\pi^+}} f_-(m_a^2)\right]\!\sum_{q=u,c,t}\! \!c^{(q)}_{sd}
\label{loopAmplitudeK+}                            
\eeq
with the coefficient 
\beq
c^{(q)}_{sd} = \frac{V_{qi}V_{qj}^*}{V_{ts}V_{td}^*} \left[3\,c_W \frac{g(x_q)}{x_t}-\frac{c_{q}\,x_q}{4\,x_t} 
                   \ln\left(\frac{f_a^2}{m_q^2}\right)\right]
\label{ALPflavorViol}
\eeq
opportunely normalized in order to factorize out all the relevant scale dependences. The penguin with the ALP emitted from the 
internal W line is included for completeness, even if in the following phenomenological analysis $c_W=0$ will be assumed. 
The dominant contribution from the penguin diagram is mostly proportional to the $c_t$ coupling. For the $K$ meson decay, with the 
charm contribution roughly accounting for $10\%$ of the total contribution.

One-loop diagrams, with the ALP emitted from the initial/final quarks can be safely neglected being suppressed by at least a factor 
$m_s^2/m_W^2 \approx 10^{-6}$ with respect to the penguin contributions, as they arise at third order in the external momenta expansion. 
Therefore, no sensitivity on the ALP--down quark couplings can emerge in the $K\to \pi \,a$ decays from one loop diagrams. 

An order of magnitude of the tree vs loop amplitude ratio is obtained from comparing Eqs.~(\ref{eq:MKLALPapprox}) and (\ref{loopAmplitudeK+}), 
giving:
\bea
R_{T/L} = \left| \frac{\mathcal{M}^T_{K^+}}{\mathcal{M}^L_{K^+}} \right| \approx 2\,\pi^2 \frac{f_K \, f_\pi}{m_t^2} 
\left|\frac{V^*_{us} \, V_{ud}}{V^*_{ts} \, V_{td}} \right| \simeq 1. \times 10^{-2} \,.
\eea
showing the expected level of suppression. Even if the tree vs loop ratio is at the per cent level, the tree level diagrams may have a 
non negligible impact in the measurement of the $K \to \pi \,a$ decays, as in principle they depend on different and less constrained, 
down quark--ALP couplings. 

Finally, the loop contribution to the $K^0_L \to \pi^0 \,a$ decay can be easily obtained from Eq.~(\ref{loopAmplitudeK+}) and reads:
\beq
{\mathcal M}^{Loop}_{K^0_L} = \frac{G_F\, m^2_t}{4\,\sqrt{2}\pi^2} \Im(V_{ts}V_{td}^*) \frac{M^2_{K^+}}{f_a}\left(1-\frac{M^2_{\pi^+}}{M^2_{K^+}}\right)
                            \left[ f_+(m_a^2) + \frac{m_a^2}{M^2_{K^+}-M^2_{\pi^+}} f_-(m_a^2)\right]\, c^{(t)}_{sd} \nn
\label{loopAmplitudeK0L}     
\eeq
being proportional to the non vanishing imaginary part of the CKM matrix.


\section{Bounds on ALP-fermion couplings}
\label{sec:pheno}

Armed with the tree--level and one--loop, charged and neutral, $K\to\pi\,a$ decays amplitudes obtained in the previous section, 
one can bound the ALP-fermion couplings using the experimental limits provided by the NA62~\cite{CortinaGil:2020fcx,CortinaGil:2021nts}, 
E949~\cite{Artamonov:2009sz,Artamonov:2008qb,Adler:2008zza} and KOTO~\cite{Ahn:2018mvc} experiments. The main assumption underlying 
the following phenomenological analysis is that the ALP lifetime is sufficiently long for escaping the detector (i.e. $\tau_a\gtrsim 
100$ ps) or alternatively the ALP is mainly decaying in a, not better specified, invisible sector. Visible ALP decays have been studied, 
for example, in \cite{DiLuzio:2020oah,Kelly:2020dda,Bauer:2017ris,Gavela:2019wzg}.

The tree--level amplitudes of Eqs.~(\ref{eq:MKALP}), (\ref{eq:MPiALP}) and (\ref{eq:MK0LALP}) depend on the ALP couplings with $s,d$ and 
$u$ quarks, while the one--loop ones reported in Eqs.~(\ref{loopAmplitudeK+}), (\ref{ALPflavorViol}) and (\ref{loopAmplitudeK0L}), are 
typically dominated by the ALP coupling with the heaviest quark running in the loop, the $t$ quark, being the $c,u$ contributions suppressed 
by the $m_{u,c}/m_t$ mass ratio barring Cabibbo enhancements. Being the focus of this paper on ALP-fermion couplings, for the rest of the 
section $c_W=0$ will be assumed. The interplay between the simultaneous presence of $c_W$ and $c_t$ has been discussed in detail in 
\cite{Gavela:2019wzg}. 

An analysis of the $K\to \pi\,a$ decay with completely general, but flavor conserving, ALP-quark couplings, would require to consider a 
five-parameters fit, $(c_u,c_d,c_c,c_s,c_t)$ beside the ALP mass $m_a$. In order to obtain meaningful information about the ALP-fermion 
couplings different simplifying assumptions have to be introduced. The phenomenological approach followed in this section will be twofold. 
First of all, in Sec.~4.1, all ALP-fermion couplings, introduced in the Lagrangian of Eq.~(\ref{eq:ALP_SMMFV}), will be assumed independent. 
Then, using the tree--level amplitudes for the s-- and t--channels, limits on $(c_u,c_s)$ and $(c_d,c_s)$ will be obtained respectively 
from the charged and neutral $K$ meson decays, setting all the other ALP-quark couplings to $0$. 
Afterwards, in Sec.~4.2, only two independent family universal ALP-fermion couplings, $c_\uparrow$ for the up quarks and $c_\downarrow$ 
for the down ones, will considered, for sake of simplicity. Under this assumption, the interplay between the tree--level and loop 
contributions to the $K\to \pi \,a$ decay will be thoroughly discussed. Limits for the universal ALP--fermion coupling $c_{a\Phi}$ can 
be then obtained straightforwardly.

\subsection{Tree--level Contributions}
\label{pheno-tree}


The tree--level amplitudes for charged and neutral $K$ decays, given by Eq.~(\ref{eq:MKALP}) and (\ref{eq:MK0LALP}) for charged and neutral 
channels respectively, in the most general case depend on four parameters: the ALP couplings to the three light quarks, $c_u, c_d$ and 
$c_s$ and the ALP mass, $m_a$. As derived in Eq.~(\ref{MpiKapprox}) the diagram with the ALP emitted by the pion contribution is strongly 
suppressed, ranging from $10^{-2}$ (in the most conservative case) to $0$ if $g_\pi=0$ is assumed. Therefore, it seems reasonable, in the 
following, to neglect the $\pi$-ALP emission diagrams, and, consequently the $K^+$ decay rate depends only on the $(c_u,c_s)$ ALP-fermion 
couplings, while the $K^0$ decay rates only on $(c_d,c_s)$ ones.

\begin{figure}[t!]
\centering
\includegraphics[scale=0.28]{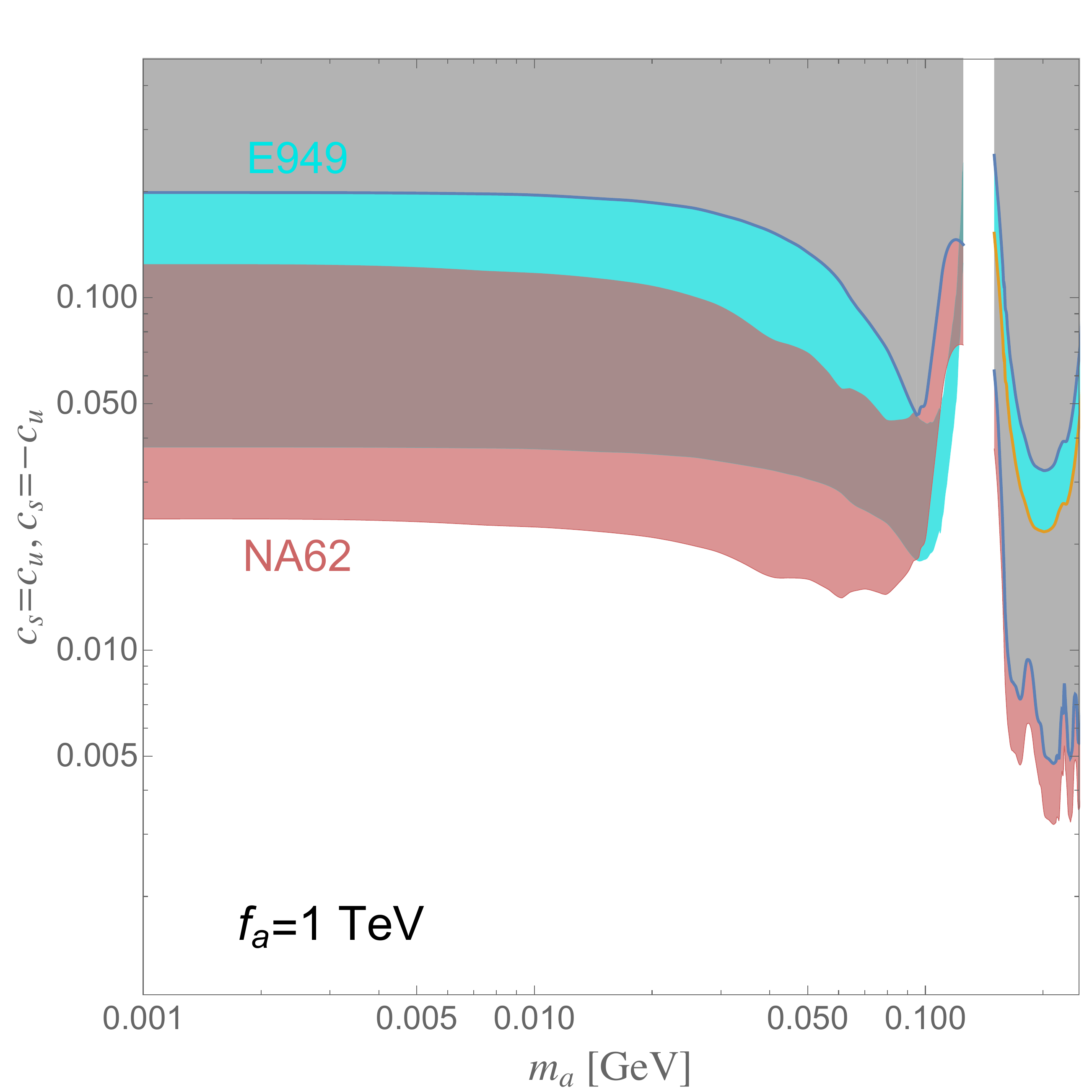}\hspace{0.5 cm}\includegraphics[scale=0.28]{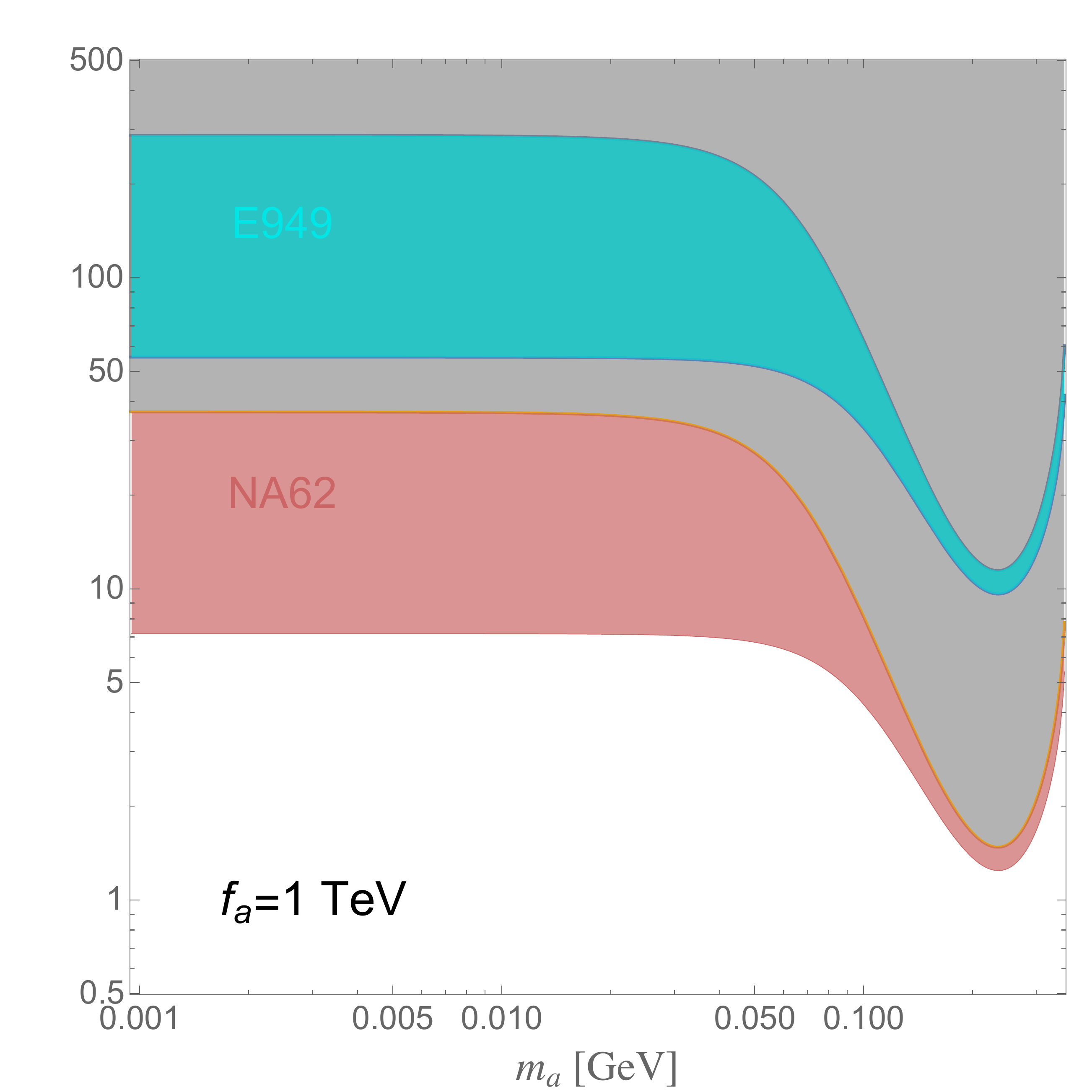}
\caption{Excluded parameter regions derived from tree-level channels for charged (left) and neutral (right) $K \to \pi a$ decays. In 
the left--plot limits are derived from NA62 (pink) and E949 (cyan) experiments. In the right--plot, bounds are obtained from present (cyan) 
and expected (pink) KOTO data.}
\label{fig:tree_plot}
\end{figure}

The left plot in Fig.~\ref{fig:tree_plot} shows the allowed regions of parameters as function of the ALP mass $m_a$ for the chosen 
reference value $f_a=1$ TeV. The shaded gray area is excluded by present experimental data. The upper and lower contours, delimiting 
the colored shaded area represent the bounds obtained setting $c_s=c_u$ and $c_s=-c_u$ respectively. The exclusive limit on $c_s\,(c_u)$ 
with $c_u\, (c_s) =0$ lies inside the colored shaded area. As noticed in Sec.~\ref{sec:hadronization}, for $c_s=c_u$ the $K^+$ decay 
rate gets suppressed by an accidental parametric cancellation, leading to a less stringent bound on the ALP--fermion couplings. The 
shaded colored area represent consequently the typical uncertainty in the bound prediction from $K^+\to \pi^+\,a$ decay rates once 
letting the couplings $(c_u,c_s)$ freely varying in the $|c_s/c_u|\leq 1$ range. The pink contours and shaded region are obtained from 
NA62 data \cite{CortinaGil:2020fcx,CortinaGil:2021nts} while the cyan ones refer to bounds obtained from the E949 \cite{Artamonov:2009sz} 
experiment. For $m_a$ values below $0.15$ GeV the two experiments provide similar results, with a slight edge in favor of NA62, bounding 
$(c_u,c_s) \lesssim 0.05$. In the $m_a > 0.15$ region, latest NA62 measurements has instead improved the sensitivity of E949 by roughly 
a factor 10, bounding $(c_u,c_s) \lesssim 0.01$. For $m_a\approx m_{\pi^0}$ both experiments loose sensitivity. No significative effects 
are obtained in this plot from the $\pi$-ALP emission diagrams, once the $c_d$ parameter is assumed to lie in the perturbative range.

A similar analysis, for $K^0_L\to \pi^0\,a$ decay is presented in the right plot of Fig.~\ref{fig:tree_plot}, where the cyan and pink regions 
are obtained using the present\cite{Ahn:2018mvc} and expected KOTO experiment data, respectively. For this plot the upper and lower contours, 
delimiting the shaded area represent the bounds obtained setting on $c_s=c_d$ and $c_s=-c_d$ respectively. KOTO experiment results much less 
sensitive to the $(c_d,c_s)$ ALP-fermion couplings, as CP violation in the tree--level processes can occurs only through the CP-mixing 
$\tilde{\epsilon}$ parameter, thus suppressing this channel by roughly a factor $10^{-3}$. Present KOTO data do not provide any real constraint, 
with the prospect that future data could reach sensitivity to the perturbativity region\footnote{Notice that KOTO experiment provides only 
mass independent limits on the $K^0_L\to \pi^0\,a$ branching ratio and consequently the loss of sensitivity in the $m_a \approx m_{\pi^0}$ 
region does not show up in the plot.}.

The results showed in Fig.~\ref{fig:tree_plot}, even if not looking flashy, represent, nonetheless, the most stringent model--independent 
bounds on light quark couplings to ALP, for an  ALP mass in the sub--GeV range, once flavour conserving, but not flavor universal 
ALP-fermion interactions, are assumed.

\subsection{Interplay between tree--level and one--loop contributions}
\label{pheno-loop}

\begin{figure}[t!]
\centering
\hspace{-3 ex}\includegraphics[scale=0.4]{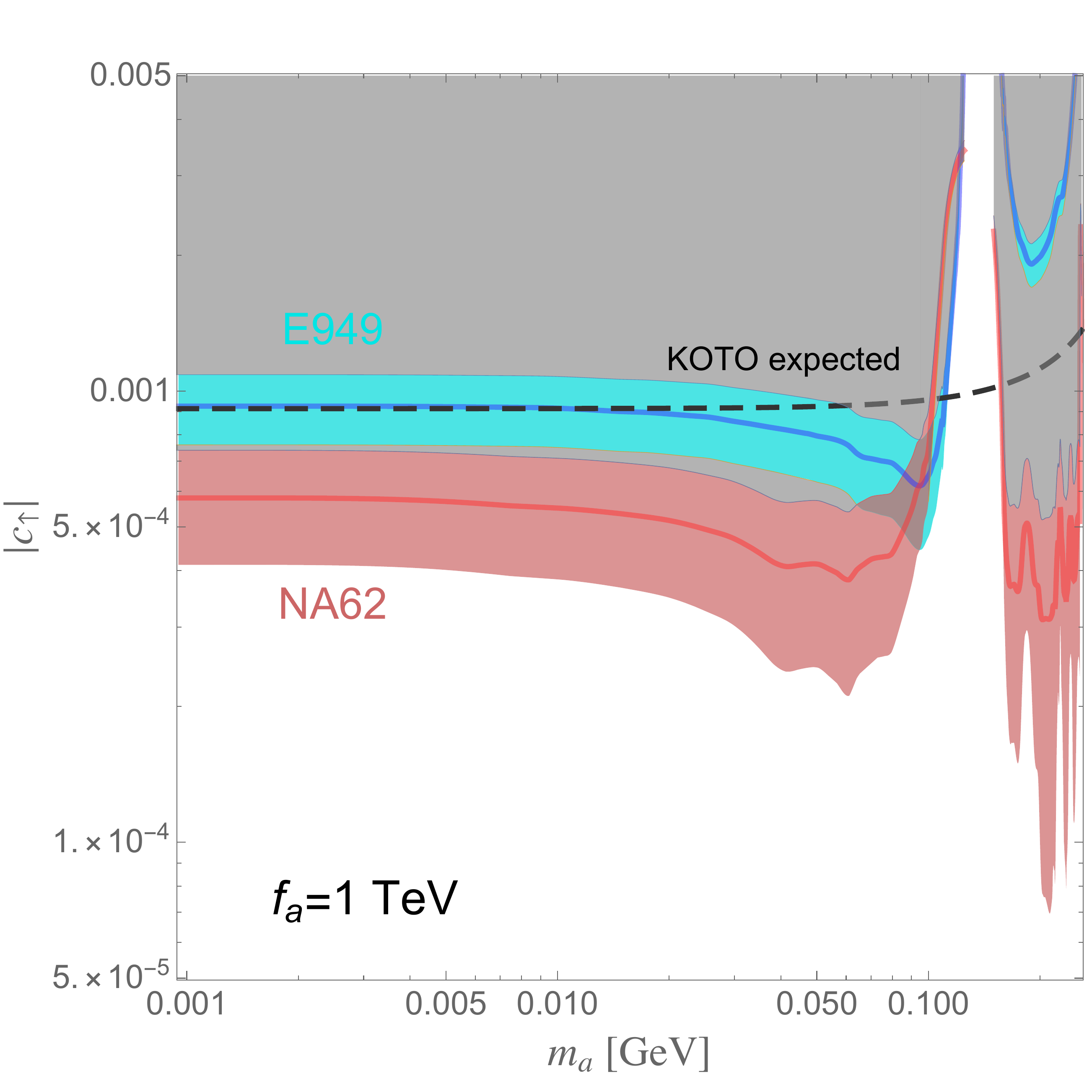}
\caption{Excluded parameter regions for an universal ALP--up quark coupling $c_\uparrow$ derived from NA62 (pink), E949 (cyan) and 
KOTO expected (dashed line) experiments.}
\label{fig:sum_plot}
\end{figure}

To constrain, simultaneously, tree--level and one--loop contributions to the $K\to \pi\,a$ decay one has to adopt simplified  frameworks. 
Following \cite{Gavela:2019wzg}, one can consider the scenario of universal ALP-quark coupling, $c_{a\Phi}$. From the analysis of 
Sec.~\ref{sec:hadronization} one easily realizes that in this scenario, the top-penguin loop contribution dominates the charged and 
neutral $K$ decay, once $c_W=0$ is assumed. The full cyan and pink lines in Fig.~\ref{fig:sum_plot}, represent the limits on $c_{a\Phi}$ 
obtained from E949 and NA62 respectively as function of the ALP mass $m_a$. The dashed gray line represents, instead, the $c_{a\Phi}$ 
limits from the expected KOTO upgrade. These results are in agreement with the bounds presented in \cite{Gavela:2019wzg} and show that 
$K$ meson decays typically constrain $c_{a\Phi}\lesssim 10^{-3}$ in the sub-GeV ALP mass range. 

\begin{figure}[t!]
\centering
\includegraphics[scale=0.28]{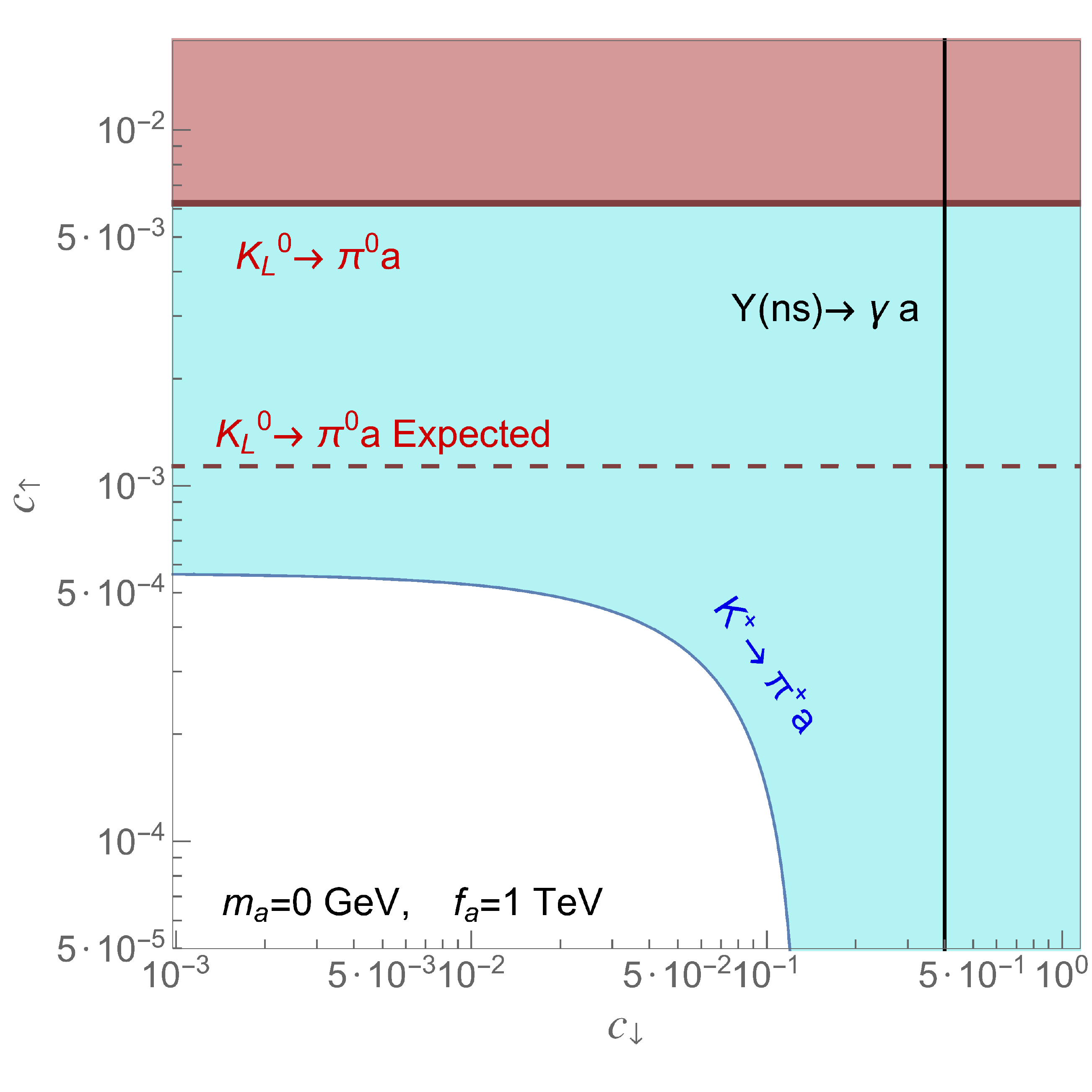}\hspace{0.5 cm}
\includegraphics[scale=0.28]{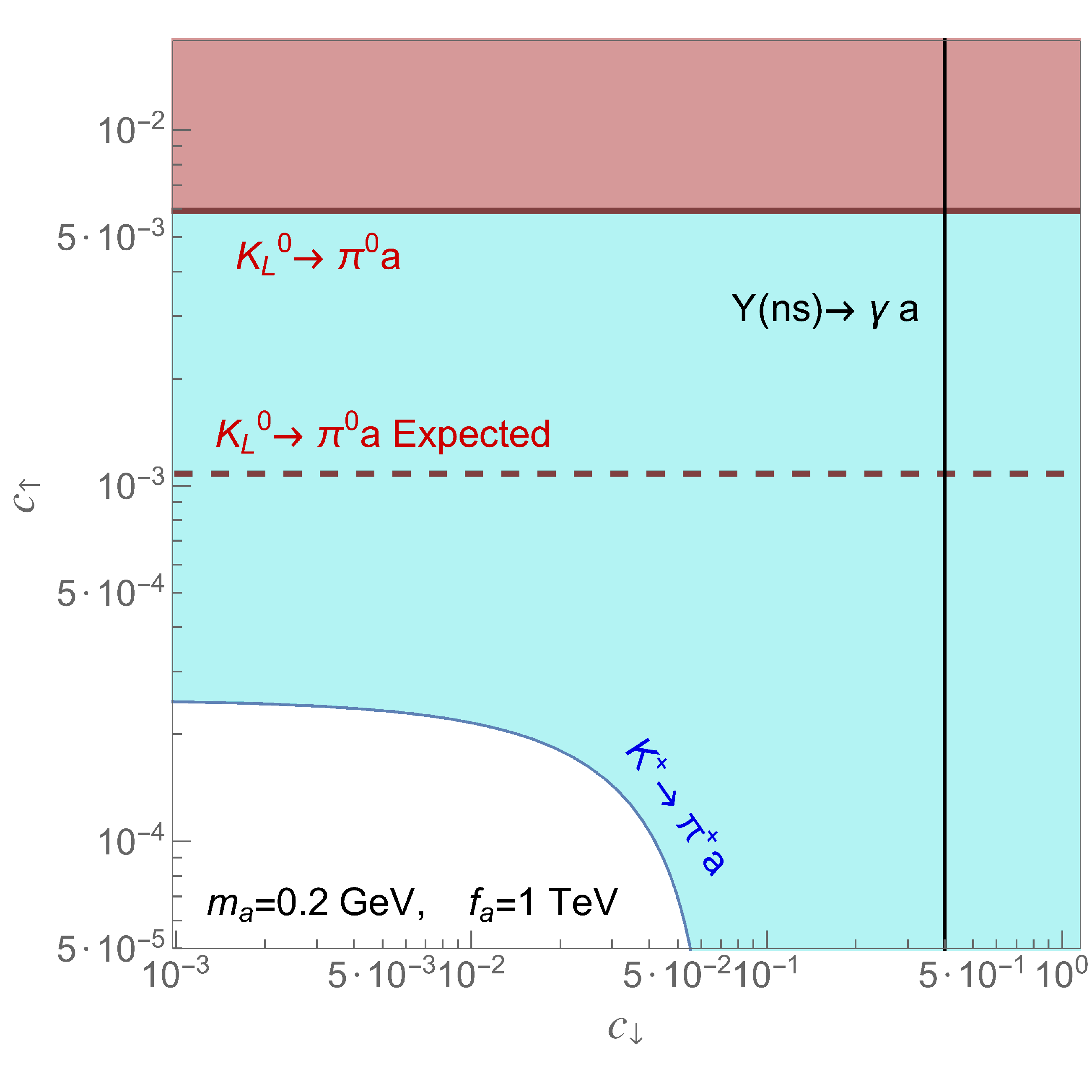} \\
\includegraphics[scale=0.28]{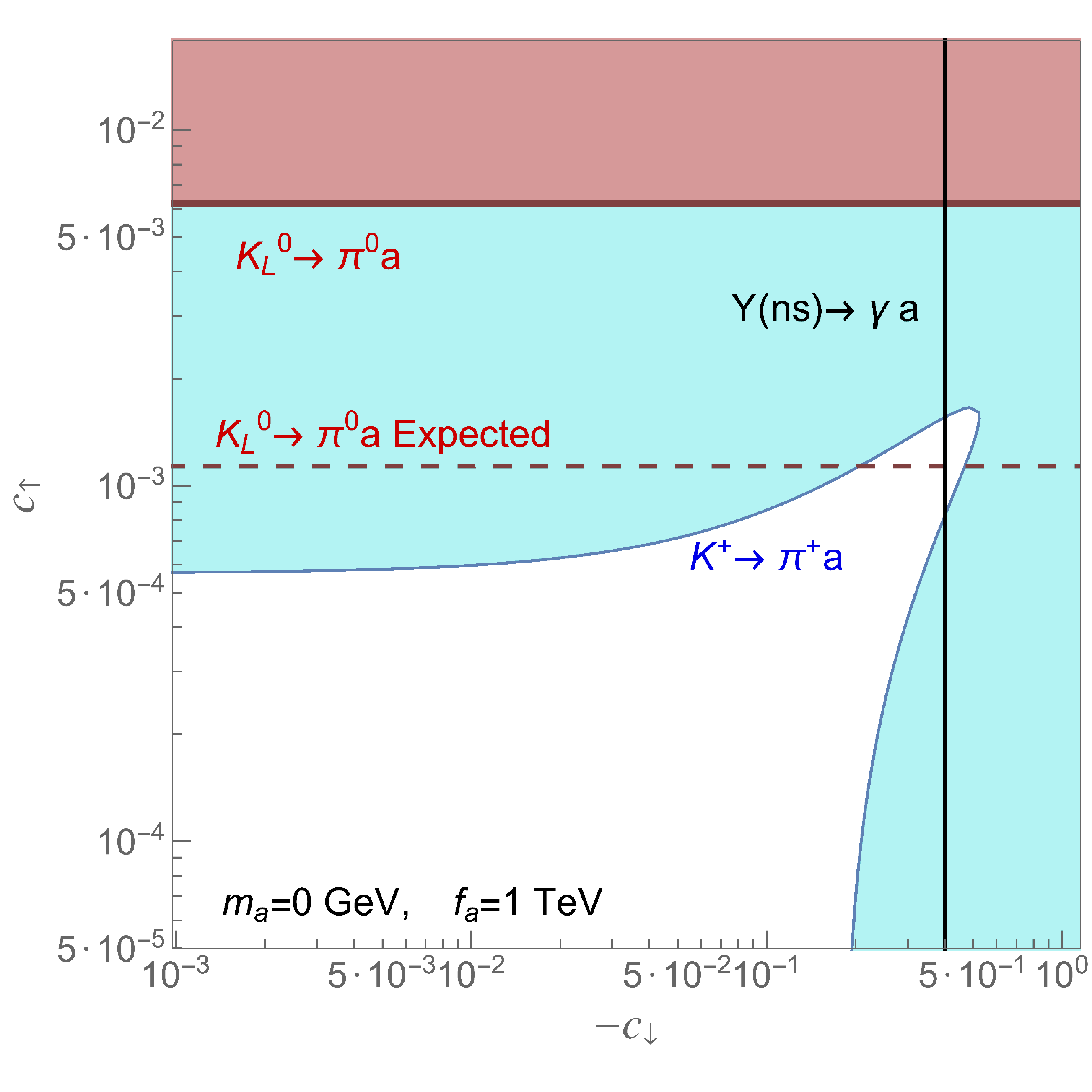}\hspace{0.5 cm}
\includegraphics[scale=0.28]{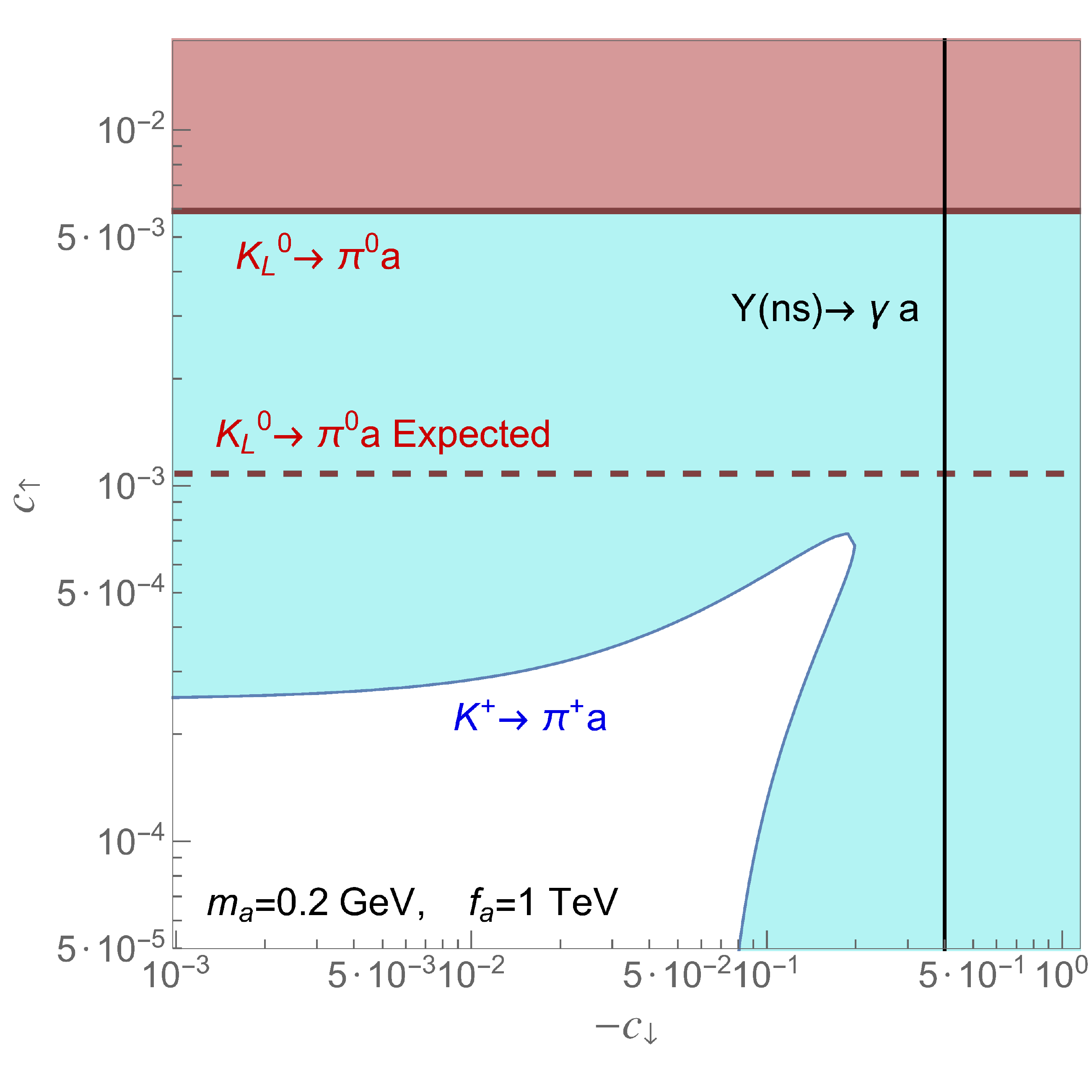} 
\caption{Excluded parameter regions for universal ALP--up and down quark couplings $c_\uparrow$ and $c_\downarrow$ derived from NA62 (cyan) 
and KOTO (pink and dashed pink line) and $Y(ns)\to \gamma \,a$ (full vertical black line) experiments. The upper plots refer to 
the $\rm{sign}(c_\downarrow) = \rm{sign}(c_\uparrow)$ case, while in the lower ones $\rm{sign}(c_\downarrow) = - \rm{sign}(c_\uparrow)$ has 
been chosen.}
\label{fig:corr_plot}
\end{figure}

In general MFV ALP frameworks, however, it may not be unconceivable to assign different, but flavor universal, PQ charges to the up and 
down quark sectors, see for example \cite{Arias-Aragon:2017eww,Merlo:2017sun,Alonso-Gonzalez:2018vpc}, that in the following will be 
denoted as $c_\uparrow$ and $c_\downarrow$, respectively. In this scenario, one--loop amplitudes only depend from $c_\uparrow$ while 
the tree--level amplitudes are practically proportional to a linear combination of $c_\uparrow$ and $c_\downarrow$, as evident for example 
in the simplified amplitudes of Eqs.~(\ref{eq:MKLALPapprox}) and (\ref{eq:MKHALPapprox}). From the tree--level analysis, summarized in 
Fig.~\ref{fig:tree_plot}, one learns that present data limit $c_\downarrow$ to be typically below $10^{-1}$. Indeed, to study the interplay 
between tree--level and one--loop the reference value $c_\downarrow=\pm 0.05$ has been chosen, somehow in the ridge of the parameters 
allowed from the previous analysis on tree--level contributions. The blue and brown shaded regions showed in Fig.~\ref{fig:sum_plot} 
represent the variability of NA62 and E979 bounds on $c_\uparrow$ once $c_\downarrow$ is let varying in the $[-0.05,0.05]$ range. 
The presence of the tree-level contribution can modify the bounds on $c_\uparrow$ extracted from penguin diagrams of roughly one order 
of magnitude, in all the $m_a$ range. The expected KOTO limits on the $K^0_L\to \pi^0\,a$ decay is reported in Fig.~\ref{fig:sum_plot} 
as a black dashed line, giving a practically constant bound $c_\uparrow \lesssim 1 \times 10^{-3} $ over all the $m_a$ range of interest, 
yet not competitive with the charged sector one. Notice that, however, the neutral $K$ decay sector does not suffer from any relevant 
interference from the tree--level processes, largely suppressed from the CP violating parameter $\tilde{\epsilon}$. A simultaneous 
measurement of the charged and neutral $K\to \pi\,a$ decays, may thus in a not too far future may give independent indications on the 
relative size of the ALP--light quark couplings.

Finally, in Fig.~\ref{fig:corr_plot}, a summary on the combined bounds on $(c_\uparrow, c_\downarrow)$ is presented for two reference values 
of the ALP mass $m_a=0$ GeV and $m_a=0.2$ GeV. For the two upper plots $\rm{sign}(c_\downarrow) = \rm{sign}(c_\uparrow)$ has been taken. In 
the lower plots, where $\rm{sign}(c_\downarrow) = - \rm{sign}(c_\uparrow)$ has been considered, a partial cancellation between one--loop and 
tree-level contributions takes place. In this second scenario, the $c_\downarrow$ constraint from the $\Upsilon(ns)$ decays at Babar 
and Belle (full vertical black line) derived by \cite{Merlo:2019anv} can contribute to close this flat direction.

\section{Conclusions}
\label{sec:conclu}

In this letter, a detailed analysis of the $K \to \pi a$ decay has been presented, in view of the recent NA62 measurement and 
the foreseen updates from the KOTO experiment. Assuming flavor and CP conserving ALP couplings with fermions, the dominant contribution 
to the $K \to \pi a$ decays arises from the penguin diagrams, manly proportional to the $c_t$ coupling. NA62 and E949 experiments 
bound $c_t$ to be smaller than $6 \times 10^{-4}$ in most of the allowed $m_a$ region, for the chosen reference value of the PQ symmetry 
breaking scale $f_a=1$ TeV. Expected KOTO results can provide comparable bounds on the $c_t$ coupling. Subdominant tree--level diagrams 
can, however, contribute to the $K \to \pi a$ decay process. The tree-level amplitudes contributing to the $K^+ \to \pi^+ a$ and 
$K^0_L \to \pi^0 a$ decays have been derived in this letter, following the Lepage--Brodsky \cite{Lepage:1980fj,Szczepaniak:1990dt} technique. 
Assuming all ALP couplings with fermions to be vanishing, but $(c_u,c_s)$, independent limits on the the light quark--ALP couplings 
$(c_u,c_s) \sim 0.05$ can be obtained in the allowed $m_a$ mass range, from NA62 and E949 experiments. KOTO experiment, conversely, will 
not provide competitive bounds on the $(c_d,c_s)$ couplings, due to the SM CP violating suppression of the $K^0_L$ channel. 

If a universal ALP coupling to fermions, $c_{a\Phi}$, is assumed, the results presented in this letter confirm and refine previous analysis 
\cite{Izaguirre:2016dfi,Dolan:2017osp,Gavela:2019wzg}, where only penguin contributions were considered. However, in general MFV ALP 
frameworks it may not be unconceivable to assign different, but flavor universal, PQ charges to the up and down quark sectors. In this case 
the simpler ''penguin dominated'' result may be significantly modified. The subdominant tree-level contribution proportional to $c_\downarrow$ 
can strongly interfere with the measurement of $c_\uparrow$, introducing an uncertainty band of roughly one order of magnitude. Similar 
analysis should be extended to all meson decays in ALP\cite{newUS}.

\section{Acknowledgements}

The authors thank O. Sumensari for useful discussions in the early stage of this paper. A.G. and S.R. acknowledge support from the European 
Union’s Horizon 2020 research and innovation programme under the Marie Sklodowska-Curiegrant agreements 690575 (RISE InvisiblesPlus) and 
674896 (ITN ELUSIVES). This project has  received funding/support from the European Union’s Horizon 2020 research and innovation programme 
under the Marie Sklodowska-Curie grant agreement No 860881-HIDDEN.


\bibliographystyle{unsrt}
\bibliography{bibliography_alf}


\end{document}